\documentclass[12pt]{article}
\usepackage{amsmath,amssymb}
\usepackage{epsfig,graphicx}
\begin{document}

\title{\textbf{Weyl Representation of the Canonical Commutation Relations Algebras in a Krein Space}
}\author{ M.\,N. Mnatsakanova$^{1}$\footnote{E-mail address:
mnatsak@theory.sinp.msu.ru}, \
 S.\,G. Salinsky$^{2}$, \
and \begin{tabular}{|c|}\hline Yu.\,S. Vernov $^{3}$\\ \hline
\end{tabular} \\
\vspace*{3mm} \\
\small $^1$Skobeltsyn Institute of Nuclear Physics, Lomonosov
Moscow State University,\\
\small  Leninskie Gory, GSP-1, 119991, Moscow, Russia \\
\small $^2$Institute for High Energy Physics,  Protvino, Russia\\
\small $^3$Institute for Nuclear Research of the Russian Academy of Sciences,
117312, Moscow, Russia}

\date{ \ }
\maketitle
\begin{abstract}
In the present article the existence of the Weyl representation for the canonical commutation relations algebras was proved in a Krein space.
\end{abstract}

\section* {Introduction \small\rm}
\mbox {} \vspace {-\baselineskip}

The canonical commutation relations (CCR) algebra lies at the foundation of a quantum mechanics. The review of basic theorems about CCR is given
in the book \cite {Put}. In one-dimensional case CCR has a well known
form:
\begin{equation}\label{1}
[\hat {p}, \hat {q}] =-i \, {I},
\end{equation}
where $ \hat {p} $ and $ \hat {q} $ are self-adjoint operators (in a
quantum mechanics they are impulse and coordinate operators accordingly). We will note,
that in the case of arbitrary, but a finite number of operators, when
the relation (\ref {1}) is substituted by more general:
\begin{equation*}
[\hat {p} _i, \hat {q} _k] = - i \, \delta _ {ik} {I}; \; 1\leq i, k\leq n,
\end{equation*}
all basic theorems for CCR can be directly generalized
\cite {Put}. So we will describe the case when the relation (\ref {1}) is fulfilled. Besides, we will describe only  irreducible representations. Among various CCR algebra representations
so called regular representations are played a special role. The Schr\"{o}dinger representation is the most known of
them. It is realized in the  space $L^{2}(-\infty, + \infty) $, i.e. in the space of functions
$f (x) $ such that $ \int ^{+\infty} _ {-\infty} \, |f (x) | ^2\,dx <
\infty $, and the scalar product of two functions $f $ and $ \phi $
is defined by the integral
$ <f, \phi> = \int ^{+\infty} _ {-\infty} f (x) \overline {\phi (x)} \, dx $. In
the given representation operators $ \hat {p} $ and $ \hat {q} $ are defined
as follows:
\begin{equation*}
\hat {q} f (q) = qf (q), \qquad \hat {p} f (q) =
-i\frac {\partial} {\partial q} f (q).
\end{equation*}
 Let us recall that there is no representation of CCR  by bounded operators \cite{Put}.  So we have to describe domains, where CCR are valid. For example for Schr\"{o}dinger representation of CCR it is done in \cite{Put}. There are various equivalent definitions
 of regular representations.  According to one of them,  representation is regular,
if it is unitary equivalent to the Schr\"{o}dinger one. The
Rellich-Dixmier`s theorem \cite{Put} says, that a representation can be unitary equivalent to Schr\"{o}dinger one, if we put very general conditions on operators $ \hat {p} $ and $ \hat {q} $.
The simplest example of non regular representations is the representation which
is defined by the same operators  as Schr\"{o}dinger  but in the space  $L^{2}(a,b)$,  the same scalar product as in Schr\"odinger one and  conditions $\varphi(a)=\varphi(b)=0$.
 Let us notice that in theories of the gauge
fields if we use a covariant gauge it is necessary to pass from Hilbert space to a space
with  indefinite  metric, because  we must introduce non-physical particles, for example, Faddeev-Popov`s ghosts in the non-Abelian gauge theory \cite{KO}. From the fact that we can not observe
non-physical particles it  immediately arises, that a scalar product, which defines observation`s probability for a corresponding particle, cannot be positive.
For an introduction of CCR representations classes which corresponds to regular representations in a
Hilbert space, it is convenient to rewrite the relations (\ref {1}) in such a form
\begin{equation}\label{2}
[{a}, {a}^ +] = 1, \; \mbox {where} \; {a} =
\frac {1} {\sqrt {2}} (\hat {q}+i\hat {p}), \; {a}^ + =
\frac {1} {\sqrt {2}} (\hat {q} -i\hat {p}).
\end{equation}
Let us consider the operator
\begin{equation}\label{N}
{N} = {a}^ + {a}.
\end{equation}
It is known that if the operator $N $ has an eigenvector:
\begin{equation}\label{4}
{N} \psi _ {\lambda} = \lambda \psi _ {\lambda},
\end{equation}
then in Hilbert space
\begin{equation*}
Sp {N} = \mathbb {N}.
\end{equation*}
Really, from a relation (\ref {4}) it follows that
\begin{equation}\label{nplusone}
{N} {a}^ + = {a}^ + ({N} +1), \qquad {N} {a} = {a} ({N}-1).
\end{equation}
Hence, with (\ref {nplusone}),
\begin{equation*}
{N} {a}^ + \psi _ {\lambda} = (\lambda+1) {a}^ + \psi _ {\lambda}, \qquad
{N} {a} \psi _ {\lambda} = (\lambda-1) {a} \psi _ {\lambda}.
\end{equation*}
The space $H $ in which CCR are realized consists of finite or convergent linear combination of operator $ {N} $ eigenvectors:
\begin{equation*}
\psi _ {\lambda+n} = ({a}^ +) ^n \psi _ {\lambda}, \qquad
\psi _ {\lambda-n} = ({a}) ^n \psi _ {\lambda}.
\end{equation*}
As in Hilbert space
\begin{equation}\label{psibeta}
<{N} \psi _ {\beta}, \psi _ {\beta}> = <{a}^ + {a} \psi _ {\beta},
\psi _ {\beta}> = (<{a} \psi _ {\beta}, {a} \psi _ {\beta}>) \geq 0
\end{equation}
then $ \beta\geq0$. From the last condition it follows that $ \lambda \in
\mathbb {N} $ and there is a vector $ \psi _ {0} $ such, that
\begin{equation}\label{11}
{a} \psi _ {0} = 0
\end{equation}
Such representations are called Fock representation. It is easy to see that
in a Schr\"{o}dinger representation $ \psi _ {0} = C \, \exp (-x^2/2) $,
where $C $  is a constant usually fixed by a normalisations condition $ <
\psi_0, \psi_0> = 1$. We will notice that all Fock representations
are unitary equivalent, and, hence, they are unitary equivalent to Schr\"{o}ding one. The fact that operators $ \hat {p} $ and $ \hat {q} $ are unbounded leads to difficulties connected with definition of domains, in which they can be set. This difficulty is eliminated if we write CCR in the Weyl form:
\begin{equation}\label{Weyl}
U (t) \, V (s) = e ^{ist} \, V (s) \, U (t),
\end{equation}
where $U (t) =e ^{it\hat {p}}, \; V (s) =e ^{is\hat {q}}, \; s, t\in\mathbb {R} $.
Boundedness of operators $U (t) $ and $V (s) $ follows from the
Stone`s theorem (see, for example \cite {Ios}), as $ \hat {p} $ and $ \hat {q} $
are self-adjoint operators. However, according to the von Neumann's theorem
and to works \cite {FGSzN}, \cite {VMS}, in Hilbert space CCR in the Weyl form
exists only for regular representations.
Let us notice that representations in the Weyl form are widely used in
quantum mechanics (see, for example, \cite {BR}).

In the present article we  consider the  representations which are defined by  condition (\ref {4}), but in a
space with indefinite metric. We prove the existence of the Weyl form of representation
for this class of representations of CCR.  Besides the case with
$ \lambda\in\mathbb {N} $ which is realized  in Hilbert space, there are two another
cases \cite {MMSV}  realized in Krein space K: \\
1. anti-Fock case: $\lambda\in\mathbb {Z _ {-}} $. \\
It is easy to see that
\begin{equation}\label{antiscpr}
<\psi _ {- k}, \psi _ {- k}> = {(- 1)}^{k - 1} {(k - 1)}!, \quad
\psi _ {- k} = a ^{k - 1} \, \psi _ {- 1}.
\end{equation}
Indeed,
\begin{equation}\label{recur}
\begin{gathered}
<\psi _ {- k}, \psi _ {- k}> = <\psi _ {- k + 1}, a ^{+} \, a \, \psi _ {- k +
1}> = (- k + 1) \, <\psi _ {- k + 1}, \psi _ {- k + 1}> = \cdots \\=
{(- 1)}^{k - 1} {(k - 1)}! <\psi _ {- 1}, \psi _ {- 1}>.
\end{gathered}
\end{equation}
We  suppose that $ <\psi _ {- 1}, \psi _ {- 1}> = 1$.  Evidently,  $ {a}^ +
\psi _ {-1} = 0.$\\
2.  $ \lambda $-case: $ \lambda = \lambda _ {0} + \mathbb {Z}, \;-1
<\lambda _ {0} <0$.

Here, by calculations  similar  to calculations for anti-Fock case, we obtain:
\begin{equation}\label{recurlamb}
<\psi _ {\lambda _ {0} + n}, \psi _ {\lambda _ {0} + n}> = (\lambda _ {0} +
n + 1) \, (\lambda _ {0} +n) \, \ldots \, (\lambda _ {0} + 1).
\end{equation}
\begin{equation}\label{recurlambneg}
<\psi _ {\lambda _ {0} - n}, \psi _ {\lambda _ {0} - n}> = (\lambda _ {0} -
n + 1) \, (\lambda _ {0} - n + 2) \, \ldots \, \lambda _ {0},
\end{equation}
where $ \psi _ {\lambda _ {0} + n} = {(a ^{+})}^{n} \, \psi _ {\lambda _ {0}},
\; \psi _ {\lambda _ {0} - n} = {a}^{n} \, \psi _ {\lambda _ {0}}, \; n \in
\, \mathbb {Z ^{+}}, \; <\psi _ {\lambda _ {0}}, \psi _ {\lambda _ {0}}> =
1$.
Let us point out that the considered space  does not contain
neutral eigenvectors of the operator ${N}$.  Let us recall that in indefinite metric space neutral vector is a vector which scalar product on the same vector is zero. In the opposite case this representation would not be irreducible.
Indeed, it is known that cyclic representation
is irreducible, if it does not depend on a choice of a cyclic vector.
If $ <\psi _ {\alpha}, \psi _ {\alpha}> =0$, then
$ <\psi _ {\alpha}, \psi> =0$, where $ \psi $  is an arbitrary vector. \\
For the proof of the given statement it is enough to consider that
the operator $N $ is self-adjoint and
$ <\psi _ {\alpha}, \psi _ {\beta}> = 0$, if  $ \alpha \neq \beta $, as
$K$ is a span of  eigenvectors of the operator $N$.
Hence,
\begin{equation*}
<P \, ({a}, {a}^ +) \, \psi _ {\alpha}, \psi> = 0,
\end{equation*}
where $P ({a}, {a}^ +) $  is an arbitrary polynomial of operators $ {a} $ and
$ {a}^ + $. It means that $ \psi _ {\alpha} $ generates a
subspace of isotropic vectors, i.e. vectors, orthogonal
to any vectors of considered space. Thus the corresponding
representation is not irreducible.

A lack of neutral eigenvectors allows to use an orthonormal  base $e_k $:
\begin {equation}\label {ika}
e_{k} = \frac {\psi_{k}}{| <\psi _ {k}, \psi _ {k}> |}.
\end{equation}
As follows from formulas (\ref {antiscpr}) - (\ref {recurlambneg}),
$ < e_{-n}, e_{-n}> \, = (-1) ^{n-1} $
for anti-Fock case, and $ < e_ {\lambda_ {0} +n}, e_{\lambda _ {0} +n}> = 1$,
$ <e_ {\lambda _ {0}-n}, e_ {\lambda_ {0}-n}> = (-1) ^{n} $ for a $ \lambda $-case.

\section* {Krein Space \small\rm}
\mbox {} \vspace {-\baselineskip}

Let us remind basic properties of a Krein space. A detailed
review is given, for example, in \cite {KREIN}, \cite {Bog}.

For enough wide class of nondegenerate spaces with indefinite metric  the following canonical expansion is known
\cite {Bog}:
\begin{equation}\label{Krein}
K = K ^{+} + K ^{-}, \; K ^{+} \perp K ^{-},
\end{equation}
where $K ^ + $  is a space with the positive metric, $K ^{-} $  is a space with the negative metric.
If $K ^{\pm} $ are closed spaces then $K $ will be a Krein space.

By definition each vector in a Krein space admits following expansion:
\begin{equation*}
x = x ^{+} + x ^{-}, \; x ^{\pm} \in K ^{\pm}.
\end{equation*}
Hence,
\begin{equation}\label{inprod}
<x, y> = <x ^{+}, y ^{+}> + <x ^{-}, y ^{-}>.
\end{equation}
In a Krein space besides an indefinite scalar product
it is possible to introduce a positive one. It is easy to see,
that
\begin{equation}\label{gilprod}
(x, y) = <x ^{+}, y ^{+}> - <x ^{-}, y ^{-}>
\end{equation}
is a positive scalar product. Indeed, for any $x \in  K,  \; x \neq 0$
$$
(x, x) = (<x ^{+}, x ^{+}> - <x ^{-}, x ^{-}>) \; > 0
$$
It is convenient to define an operator $ {J} $ of the canonical symmetry:
\begin{equation}\label{cansym}
{J} \, (x ^{+} + x ^{-}) = x ^{+} - x ^{-}
\end{equation}
Then positive and indefinite scalar products are connected as follows:
\begin{equation}\label{connect}
<x, y> = (x, {J} y), \qquad (x, y) = <x, {J} y>.
\end{equation}
It is easy to check following properties of an operator $ {J} $: \\
1. $ {J}^2 = 1, $ \\
2. $ {J} = {J}^{-1} $. \\
3. $ {J} $  is a self-adjoint operator: $ {J} = {J}^{+} = {J}^ * $
for both, indefinite and positive, scalar products accordingly.

It is possible to show how operators $A^{+} $ and $A^{*} $,  which  are adjoint ones according to indefinite and positive scalar product correspondingly,  are connected in a Krein space. Indeed,
\begin {gather*}
<{A} x, y> = ({A} x, {J} y) = (x, {A}^ * {J} y) = <x, {J} {A}^ * {J} y>.
\end {gather*}
Hence,

\begin{equation}\label{88}
{A}^{+} = {J} {A}^ * {J}.
\end{equation}

\section* {Proof of the Naimark theorem
for regular CCR representations \small\rm}
\mbox {} \vspace {-\baselineskip}

The existence of the Weyl form of CCR in a Krein space is connected
with the Naimark theorem \cite {Naim}, being generalisation of the
Stone theorem in a space with  indefinite metric. \\
{\bf Naimark theorem}. \quad{\it
Let $B$ be a self-adjoint operator in a Krein space.
If $B$ satisfies  for all $n \in \mathbb Z$  with sufficiently large $ |n|$ and some $ M > 0, \beta > 0$:\\
1. $ (I - i \, n ^{-1} \, {B}) ^{-1} $  is a bounded operator; \\
2. $ \|(I - i \, n ^{-1} \, {B}) ^{-m} \| \leq M{ (1 - | n ^{-1} | \, \beta)}^{-m}, \; m=1,2,..., $ \\
then  $A = i \, B$  is the generator of some one-parameter group  $U (t) $ of unitary operators in the Krein space satisfying the following conditions:
\begin{enumerate}
    \item [a)]  $U\,(t_{1} + t_{2}) = U\,(t_{1}) U\,(t_{2}) $;
    \item [b)] $U\,(t)\,x$ is a continuous in the norm function of $t$ for any $x$.
\end{enumerate}}
As the  space in question is a linear span  of defined in (\ref{ika}) eigenvectors $e_{-k}$ of an operator  ${N}$
we  obtain the dense domain consisting of vectors
\begin{equation}\label{881}
\Psi^{l} = \sum ^{l} _ {k=1} {C_k \, e _ {-k}}.
\end{equation}
 As it was noted earlier,
the operator ${N}$ has no  neutral  eigenvectors.

Let us notice that $ {a} \, e _ {-k} = b _ {k} \, e _ {-k-1} $. On the other hand,
$$
<{a} \, e _ {-k}, {a} \, e _ {-k}> = - k \, <e _ {-k}, e _ {-k}>,
$$
using a condition $ <e_{-k}, e_{-k}> \, =  (-1) ^{k-1} $, we obtain that
$b_{k} = \sqrt {k} $. It is similarly shown that ${a}^{+} \, e_{-k} =
d _ {k} \, e _ {-k+1} $, and $d _ {k} = \sqrt {k-1} $. It is obvious that
\begin{equation}\label{connec}
d _ {k} = b _ {k-1} = \sqrt {k - 1}.
\end{equation}
It is easy to prove that
\begin{equation}\label{anti}
\{{a}, {J} \} = \{{a}^ +, {J} \} =0; \; \{x, y \} = x \, y + y \, x.
\end{equation}
According to (\ref {88}) and (\ref {anti}),
\begin{equation}\label{conj}
{a}^ + = - {a}^*
\end{equation}
and
$$
{I} -i \,  \frac {1}{n\,\sqrt{2}} ({a} + {a}^ +) =  {I} - i \, \frac {1}{n\,\sqrt{2}} ({a} - {a}^ *).
$$
We denote:
\begin{equation}\label{denote}
A = {I} - i \,  \frac {1}{n\,\sqrt {2}} ({a}  - {a}^ *) .
\end{equation}
 Let us consider decomposition
$$
 K = K_{1} + K_{2},
$$
where $K_{1} $ is a span of real vectors and $K_{2}$ is a span of imaginary vectors.
So
$$
\Psi^{l} = \Psi_{1}^{l} + \Psi_{2}^{l},
$$
where
$$
\Psi_{1}^{l} = \sum \limits_{1}^{l}C_{k}e_{-k};  \quad \Psi_{2}^{l} = i\,\sum \limits_{1}^{l}C_{k}^{'}e_{-k}, \quad  C_{k}, C_{k}^{'} \in \mathcal R.
$$
So every vector in $K$ is a sum of vectors in $K_{1}$ and $K_{2}$.
First we prove that Naimark theorem is valid in space $K_{1}$, that is for vectors $\Psi_{1}^{l}$.
Let us prove the Condition 1 for the operator $ \hat {q} = \frac {1}{\sqrt{2}} \, (a + a^ +) $  (see (\ref {4})).
Let us show that there are such $ \alpha > 0$ that
\begin{equation}\label{99}
\left \|A\, \psi \right\|> \alpha \,||\psi||, \quad \mbox{where} \quad \psi \equiv\Psi_{1}^{l}.
\end{equation}
First, we give the proof for anti-Fock case.

Let us prove  inequality (\ref {99}) for vectors $ \psi $
(\ref {881}) for $ \alpha = 1$.
\begin{equation} \label{1000}
\begin {gathered}
\left \|A\, \psi\right \| ^ 2 = \\
=
\left (\left ({I}-i \,  \frac {1}{n\,\sqrt {2}} ({a}  - {a}^ *) \right) \psi,
\left ({I} - i \, \frac {1}{n\,\sqrt {2}}({a}   - {a}^ *) \right) \psi\right) = \\
= {\| \psi \|}^{2} +
{\left \|-i \,  \frac {1}{n\,\sqrt {2}} ({a}   - {a}^ *) \psi \right \|}^{2}
 + \left  (-  \frac{i\,\sqrt{2}}{n} ({a}   - {a}^ *) \, \psi, \psi\right ).
\end {gathered}
\end{equation}
So
\begin{equation} \label {33}
\begin{gathered}
 \frac{-i\,\sqrt{2}}{n}\left( ({a} - {a}^*) \, \psi,  \psi\right) =
\frac{2\,\sqrt{2}}{n} Im \, \left(({a} - {a}^*) \, \psi, \psi\right).
\end{gathered}
\end{equation}
In order to make clear the last term in eq. (\ref{1000}) let us notice that
$$
\left(\psi, {\frac{-i}{n\,\sqrt{2}}}(a - a^*)\psi\right) =
\frac{i}{n\,\sqrt{2}}\left((a^{*} - a)\,\psi, \psi\right) =  \frac{- i}{n\,\sqrt{2}}\left((a - a^{*})\,\psi, \psi\right) .
$$
Let us recall that in $K_{1}$ coefficients  $C_k$  in (\ref{881}) are real.  Thus   $\psi$ and, in accordance with (\ref{connec}), ${a}\psi$ and ${a}^ * \, \psi $ are real vectors and the latter term in eq.(\ref{1000}) is equal to zero.

Let us show that the Naimark theorem is true in the $ \lambda $-case. For this purpose we notice that the vector (\ref {881}) is substituted by
\begin{equation*}
\psi = \sum ^{l} _ {k =-l} {C_k e _ {\lambda+k}}; \qquad e_ {\lambda+k} =
\frac {\psi _ {\lambda+k}} {| <\psi _ {\lambda+k}, \psi _ {\lambda+k}> |}.
\end{equation*}
Thus  the Condition 1 of the Naimark theorem is  proved for both cases.

The Condition 2 of the theorem directly follows from the fact that $ \alpha \geq 1$ (see (\ref {99})). Really, as
$ \, {A} ^{-1} $,   is a bounded operator, then
\begin{equation*}
\begin {gathered}
\left \|  {A^{-m}} \right \|  \leq
\left \|  {A} ^{-1} \right \| ^ m.
\end {gathered}
\end{equation*}
As $ \alpha \geq 1$, then there always exists a number $\beta > 0$, such that
\begin{equation*}
\begin {gathered}
\left \|  \, {A}^{- 1} \right \| \leq
{(1 - n^{-1} \, \beta)}^{-1}.
\end {gathered}
\end{equation*}
It is easy to check by a similar way  Conditions 1 and 2 for the operator $ \hat {p} $.
For space $K_{2}$ we have the same proof. Only we have to notice that in accordance with   (\ref{connec}) as  $\Psi_{2}^{l}$ is an imaginary vector, then $a\, \Psi_{2}^{l}$ and  $a^{*}\, \Psi_{2}^{l}$ are imaginary vectors as well.
 We can always  define $U(t),\:  t=t_{1} +t_{2}$
as follows: $U(t)=U(t_{1})U(t_{2})$, where $U(t_{1})$ and $U(t_{2})$ belong to $K_{1}$ and $K_{2}$ correspondingly.
 In order to proof  Naimark theorem for the vectors $ \Psi^{l}$ it is sufficient to notice that
$$
\| U\,(t)\,( \Psi_{1}^{l} +  \Psi_{2}^{l}) \| \leq \|U\,(t)\, \Psi_{1}^{l}\|  +   \| U\,(t)\,\Psi_{2}^{l}) \|.$$

As $U(t)$ on $K_{1}$ and $U(t)$ on $K_{2}$ are bounded operators, then $ U(t)$ on $K$ is a bounded operator as well.

\section*{Existence of the Weyl form of CCR in dense domain of the considered space $ K $
\small\rm} \mbox {} \vspace {-\baselineskip}

Now let us notice that if  $D$  is a dense domain, consisting
of analytical vectors for operators $p $ and $q $, then CCR in the Weyl form exists in $D $ for Hilbert space (the proof see in \cite {VMS}).As $U(t)$ is a bounded operator then CCR in the Weyl form exist in the whole space in question.
 In this paper we extend this result for a Krein space.
The proof that from the existence of analytical vectors in some domain it follows that in this domain there exist CCR in the Weyl form  for Krein space is similar to the corresponding proof for Hilbert space.
A definition of an analytical vector is given in \cite {Nels}. Analytical vectors are defined
in any Banach space, hence, for a Krein space too. \\
{ \bf  Definition} \quad{\it
Let $ {A} $ be a linear operator in a Krein space $K$. A vector
$ \xi \in K $ is called an analytical vector for the operator ${A}$,
if $ \xi $ belongs to a domain of definition of operators
$ {A^k}$ for any  $k \in \, \mathbb {N}$ and for each $t > 0$ the series
\begin{equation*}
\sum ^{\infty} _ {k=0} \frac {t^k} {k!} \, \| {A}^k \, \xi  \|
\end{equation*}
converges.} \\
For Fock representations it has been shown that analytical vectors exist \cite{VMS}.
Results of this paper can be generalized on anti-Fock and $ \lambda $-cases.

Let us consider a vector
\begin{equation*}
\psi = \sum^{m+n} _m C_l \, \psi _ {- l}.
\end{equation*}
According to (\ref {antiscpr}), $ \| \psi _ {- l} \| = \sqrt {(l - 1)!} $.
Hence,
\begin{equation*}
 \| \hat {q} \, \psi \| \leq \sum ^{m+n} _m \, \| \hat {q} \, C_l \, \psi _ {- l} \| \leq \sqrt {2} \,n C \, \sqrt {(m + n - 1)!},
\end{equation*}
where  $C = \max |C_l |, \; m \leq l \leq m + n $. We have used that $ \| a\, \psi _ {- l} \|> \| a ^{+} \psi _ {- l} \| $.

For a similar estimation with $ \| {\hat {q}}^{k} \, \psi \| $,
let us notice that in the sum of norms the greatest value has $ \| {a}^k \, \psi \| $.
\begin{gather}\label{19}
\left \| \frac {1} {\sqrt {2}} \, {(a + {a}^ +)}^{k} \, \psi \right \| \leq
2 ^{\frac {k} {2}} \, \left \| a ^{k} \, \psi \right \| \leq \tilde{C}\,n{\sqrt {2} \,}^{k} \, \sqrt{(m + n +k) !},
\end{gather}
where $\tilde{C} =  \max |C_l |, \; m \leq l \leq m + n + k$.
The inequality (\ref {19}) is written taking into account its possible
applications in a $ \lambda $-case.

According to inequality (\ref{19}),
\begin{equation}\label{48}
\sum ^{\infty} _ {k=0} \frac {t^k} {k!} \, \| {q}^k \, \psi \| \leq
\sum ^{\infty} _ {k=0} \frac {t^k} {\sqrt {k!}} \,\tilde{C^{'}}\,n {(\sqrt {2} \,)}^{k} \,{(m + n
+k)}^{m + n}.
\end{equation}
Let us notice that for every vector in  space under consideration $\sum^{\infty} _ {l=1}{|C_{l}|}^{2} < \infty$ and so $C_{l} \to 0$ if $l \to \infty$.
Convergence of the series (\ref {48}) at arbitrary $t $ and any finite
$n $ and $m $ can be easily established\footnote {In the formula
(20) of our paper \cite {VMS} there was a misprint, which is not influencing the
outcomes of the work. The correct formula coincides with the formula
(\ref{48}) in the present paper.}. It is enough to consider the Stirling formula: $k! \simeq
{\sqrt {2 \, \pi \, k}} {\left (\frac {k} {e} \right)}^{k} $.

Thus, it is proved that a vector $ \psi $ is an analytical vector
for the operator $ \hat {q} $ in dense domain $D $. The proof
that a vector $ \psi $  is an analytical vector for the operator
$ \hat {p} $ can be made by a similar way.
 As $u(t)$ is a bounded operator in the space under consideration CCR in the Weyl form exist on the whole Krein space.
So, results of the work \cite {VMS} and the Naimark`s theorem allows
us to extend the Weyl relations to a full Krein space.

The proof for a $ \lambda $-case is reduced to proofs for
Fock and anti-Fock cases. Really, if we divide as earlier,
$\psi = \psi_{\lambda_{0} + n} + \psi_{\lambda_{0} - n}$, we can see with (\ref {recurlamb}) and (\ref {recurlambneg}),
\begin{equation*}
| <\psi _ {\lambda _ {0} + n}, \psi _ {\lambda _ {0} + n}> | <n!,
\end{equation*}
\begin{equation*}
| <\psi _ {\lambda _ {0} - n}, \psi _ {\lambda _ {0} - n}> | <n!.
\end{equation*}
Thus, a $ \lambda $-case is reduced to the sum of Fock and
anti-Fock cases.

\section*{Comparison of the Weyl  form of CCR  in Hilbert and Krein spaces}

To compare the Weyl form of CCR  in Hilbert and Krein spaces first let us notice  that for Anti-Fock
 case there exists analogue of the Weyl form of CCR \cite{VMS}.
For reader convenience we give the short proof of this assertion.
First let us notice that if $a = b^{+}$,   $b = a^{+}$, then anti-Fock representation is written as follows:
\begin{equation}\label{bbkrest}
[b,  b^{+}] = - 1.
\end{equation}
Introducing operator $\tilde{N}$  and the set of its eigenvectors $e_{\tilde{N}} =e_{- N - 1} $,  we see that
\begin{equation*}
\tilde{N}\,e_{\tilde{N}} = \tilde{N}, \quad Sp \tilde{N} =\mathbb{N}
\end{equation*}
In order to come to analogue of the Weyl  form of CCR first notice that it is easy to show that
$\{J\, b\} = \{J\, b^{+}\} = 0$, where $\{x\, y\} = xy +yx$ (see (\ref{88})).

Now let us consider the pair of operators $b$ and  $b^{*}$,  where $b^{*} = J\, b^{+}\,J = - b^{+}$ in accordance with eq. (\ref{88}).
Evidently
\begin{equation}\label{bbstar}
[b,  b^{*}] = 1.
\end{equation}
This equation is a usual Fock representation of CCR and thus operators  $U (t) =e ^{it\hat {p}}$ and  $V (s) =e ^{is\hat {q}}$, where $s, t\in\mathbb {R} $ satisfy the Weyl form of CCR (\ref{Weyl}).
Returning to the operators  $b$ and  $b^{+}$,   we come to the analogue of the Weyl form of CCR for anti-Fock representation.
Let us turn Hilbert
space to the Krein space. We  remind the well-known procedure. First
in Hilbert space new, in general, indefinite scalar
product  ${(x, y)}_{A}\equiv (x, Ay)$, where  operator $A$ is  self-adjoint, is  introduced.
Choosing as $A$ operator $J$, which is simultaneously self-adjoint and
unitary and using the projective operators $\Pi_{\pm} = 1/2\,{(I \pm J)}$, we come to the
Krein space $K$ (\ref{Krein}),  where  $K_{\pm} = \Pi_{\pm} \mathcal H$.

It is easy to see that  we can come from Fock  representation to the anti-Fock one repeating in the opposite order the way from anti-Fock to the Fock representation (see eqs.(\ref{bbkrest}) and (\ref{bbstar})).

\section* {Conclusion \small\rm}
\mbox {} \vspace {-\baselineskip}

An existence of the Weyl form of CCR is proved for the regular representations in a space with indefinite metric.

\medskip

\noindent {\bf Acknowledgements}.
We thank Professor  A.D. Baranov for very useful mathematical discussions.


\begin{thebibliography}{99}
\bibitem{Put}
Putnam C.R., Commutation properties of Hilbert space
operators and related topics.  Springer-Verlag. Berlin;
Heidelberg; New York. 1967.

\bibitem{KO}
Kugo T. and Ojima I., Suppl. Prog. Theor. Phys. 1979.
{\bf 66}. P. 1.

\bibitem{Ios}
Iosida K., Functional analysis. 6th ed., Springer-Verlag,
Berlin-New York, 1980.

\bibitem{FGSzN}
Foias C., Geh\'{e}r  L. and Sz.-Nagy B., Acta Sci. Math. (Szeged)   1960.
{\bf 21}. P. 78.

\bibitem{VMS}
Vernov Yu S., Mnatsakanova M.N., Salynskii S.G., PhysPNLt.   2012.
{\bf 3}. P. 213.

\bibitem{MMSV}
Mnatsakanova M.,  Morchio G., Strocchi F. and Vernov Yu.,
Jour. Math. Phys. 1998.  {\bf 39}. P. 2969.

\bibitem{BR}
Bratteli O. and Robinson D.W., Operator algebras and
quantum statistical mechanics. {\bf V. 2}. Springer-Verlag.
Berlin-Heidelberg-New York. 1979.

\bibitem{KREIN}
Krein M. G., Amer. Math. Soc. Transl. 1970.  {\bf 93}. P. 103.

\bibitem{Bog}
Bognar J., Indefinite inner product spaces.
Springer-Verlag. Berlin-Heidelberg-New York. 1974.

\bibitem{Naim}
Naimark, Doklady akademii nauk SSSR. 1966. {\bf 170}. P. 1259.

\bibitem{Nels}
Nelson E., Ann. Math. 1959. {\bf 70}. P. 572.

\end{thebibliography}
\end {document}